\documentclass[fleqn,12pt,twoside]{article}
\usepackage{espcrc1}

\usepackage{graphicx}
\usepackage[figuresright]{rotating}

\newcommand{\AmS}{{\protect\the\textfont2
  A\kern-.1667em\lower.5ex\hbox{M}\kern-.125emS}}

\title{High-$p_T$ $\pi^0$ suppression in Au+Au collisions at $\sqrt{s_{\mbox{\tiny{\it{NN}}}}}$ = 200 GeV.}

\author{David d'Enterria\address{SUBATECH, 4 rue A. Kastler, BP 20722, 44307 Nantes Cedex 3, Bretagne, France}\thanks{Supported by 5th European Union TMR Programme (Marie-Curie Fellowship HPMF-CT-1999-00311).} % is acknowledged.} 
	for the PHENIX Collaboration\thanks{For the full PHENIX author list and acknowledgements, 
see Appendix "Collaborations" of this volume.}}
       
\begin{document}

% typeset front matter
\maketitle

\begin{abstract}
%\hspace{-5mm}
Neutral pions with $p_T =$ 1 - 8 GeV/c have been measured for 9 centrality classes in
Au+Au collisions at $\sqrt{s_{\mbox{\tiny{\it{NN}}}}}$ = 200 GeV by the PHENIX experiment
at RHIC. The $\pi^0$ multiplicity in central reactions is significantly below the binary collision 
scaled yields from both peripheral Au+Au and $pp$ reactions. The observed suppression sets 
in for the 50-70\% centrality class and increases with $p_T$ and centrality. For the most 
central bin, the deficit amounts to a factor $\sim$2.5 at $p_T\sim$ 2 GeV/c gradually 
increasing to a factor $\sim$6 at $p_T\sim$ 8 GeV/c.
\end{abstract}

\section{Introduction}
One of the primary goals of the PHENIX experiment at BNL RHIC is the 
measurement of hadron spectra out to large transverse momentum ($p_T$) in high-energy heavy-ion 
(HI) collisions. High-$p_T$ particles result from the fragmentation of quarks and gluons produced 
in large $Q^2$ parton-parton scattering processes during the earliest stages of a HI collision 
and provide direct signatures of the partonic phase of the reaction. Interestingly, hard scattering 
multiplicities can be quantitatively compared (after scaling with the number of binary nucleon-nucleon, 
$NN$, inelastic collisions $N_{coll}$) to: (i) baseline ``vacuum'' ($pp$) and ``cold medium'' 
($pA$) data, and (ii) perturbative QCD predictions. %through factorization at high momentum transfers
In both cases, any departure (deficit or excess) from the ``expected'' results provides %precious 
information on the strongly interacting hot and dense medium created in the reaction.
One of the most intriguing results from the first RHIC run was the suppressed yield 
of moderately high-$p_T$ $\pi^0$ ($p_T\approx$ 1.5 - 4.0 GeV/c) in central
Au+Au collisions at $\sqrt{s_{\mbox{\tiny{\it{NN}}}}}$ = 130 GeV
with respect to the scaled $pp$ extrapolation \cite{enterria:ppg003}. Such a suppression is in 
qualitative agreement with theoretical predictions of parton energy loss effects 
(``jet quenching'') in an opaque medium, the amount of energy loss being 
proportional to the reached gluon density \cite{enterria:ina02}.

\section{Experimental setup and data analysis}

During the 2001-2002 RHIC run, the full PHENIX Electromagnetic Calorimeter EMCal was 
instrumented providing a solid angle coverage of approximately $\Delta\eta = 0.7$
and $\Delta \phi = \pi$. The integrated luminosity of $\sqrt{s_{\mbox{\tiny{\it{NN}}}}}$ = 200 
GeV Au+Au collisions was about 25 $\mu$b$^{-1}$ and the collected neutral pion statistics 
was a factor of $\sim 10^{2}$ larger than in Run-1, allowing the exploration of the $\pi^0$ 
spectra up to much larger values in $p_T$ and for finer centrality classes. %Additionally, 
PHENIX also measured the $\pi^0$ spectrum in $pp$ collisions at the same $\sqrt{s}$ \cite{enterria:hisa02}. 
The combination of full acceptance, high statistics, and the measurement of $pp$ data in the same detector 
permits a very precise study of the high-$p_T$ $\pi^0$ %production and 
suppression at 200 GeV.

The data presented in this analysis correspond to pions from 30$M$ $AuAu$ ``minimum bias'' events 
with vertex position $|z|<$ 30 cm. The PHENIX EMCal consists of 4 lead-scintillator (PbSc) sectors 
(2592 towers per sector with 5.25$\times$5.25$\times$37.0 cm$^3$ size) in the west central arm plus 
2 more sectors in the east one, and 2 lead-glass (PbGl) sectors (4608 towers of 4.0$\times$4.0$\times$40.0 cm$^3$) 
in the east arm. 
%Each sector covers roughly $|\eta|=0.35$ units of pseudorapidity at mid-rapidity, and $\Delta\phi=22.5^\circ$ in azimuth. 
The  large radial distance of the calorimeter to the interaction region, of $\sim$5 m, 
keeps the detector occupancy reasonably low ($<$15\%) even in the highest multiplicity events of HI collisions at RHIC. 
Event centrality is determined by correlating the charge and energy measured in two global 
detector systems: the Beam-Beam Counters and the Zero Degree Calorimeters.
Neutral pions are reconstructed through an invariant mass analysis of $\gamma$ pairs 
detected in the 8 active EMCal sectors. The raw $\pi^0$ spectra are then corrected for geometric
acceptance and (multiplicity-dependent) reconstruction and efficiency losses. 
The corrections were determined embedding 2$M$ simulated $\pi^0$ into real events. 
The final systematic errors in the fully corrected spectra are of the order of $\pm$25\%. 
Two independent analyses were carried out for the PbSc (shown here) and PbGl calorimeters 
yielding consistent results within the overall systematic error.
%The results presented here are those of the PbSc calorimeter.

\section{Results}

\begin{figure}[htb]
\begin{minipage}[t]{75mm}
  \includegraphics[scale=0.35]{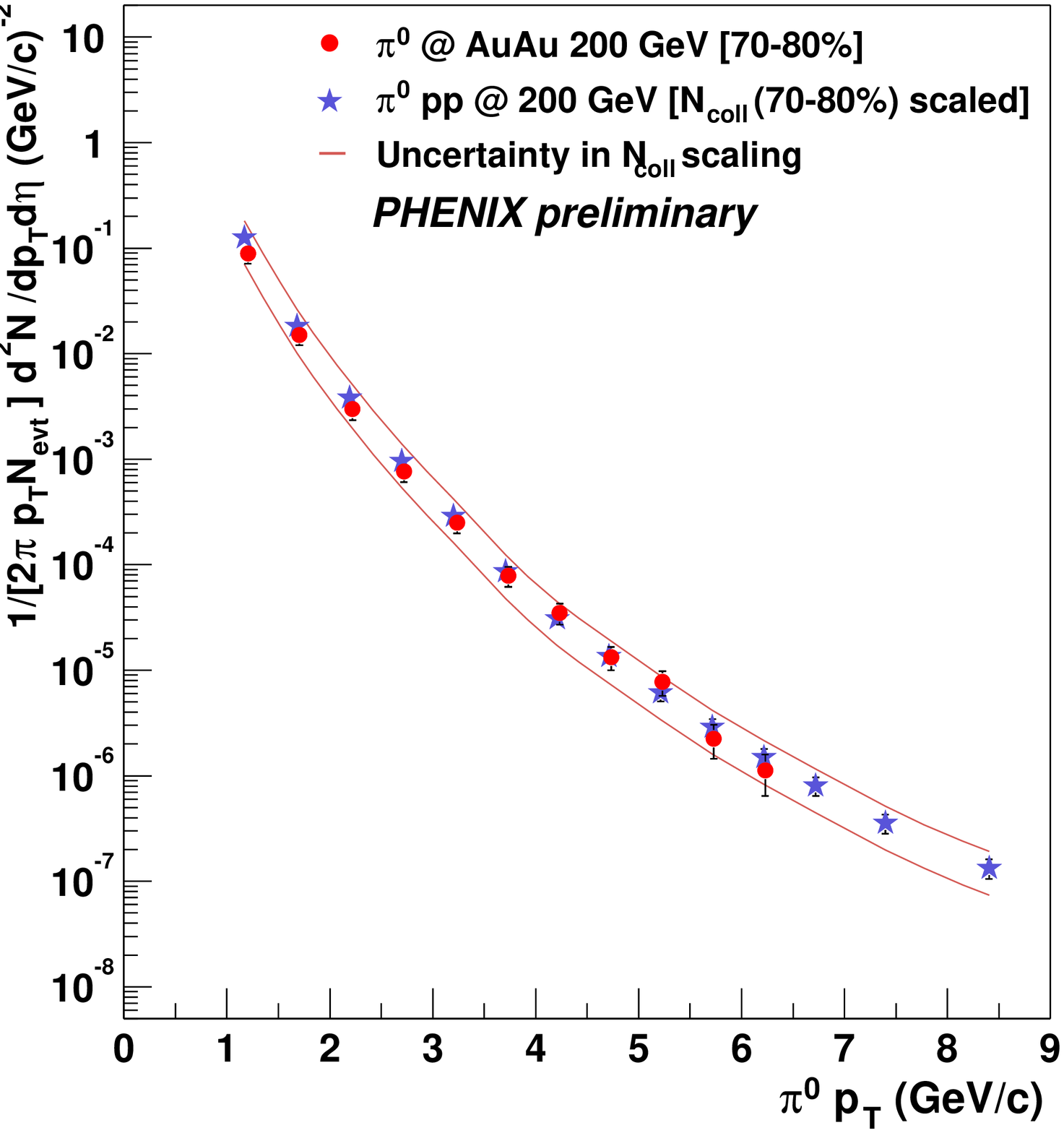}
\end{minipage}
%\hspace{\fill}
\begin{minipage}[t]{75mm}
  \includegraphics[scale=0.35]{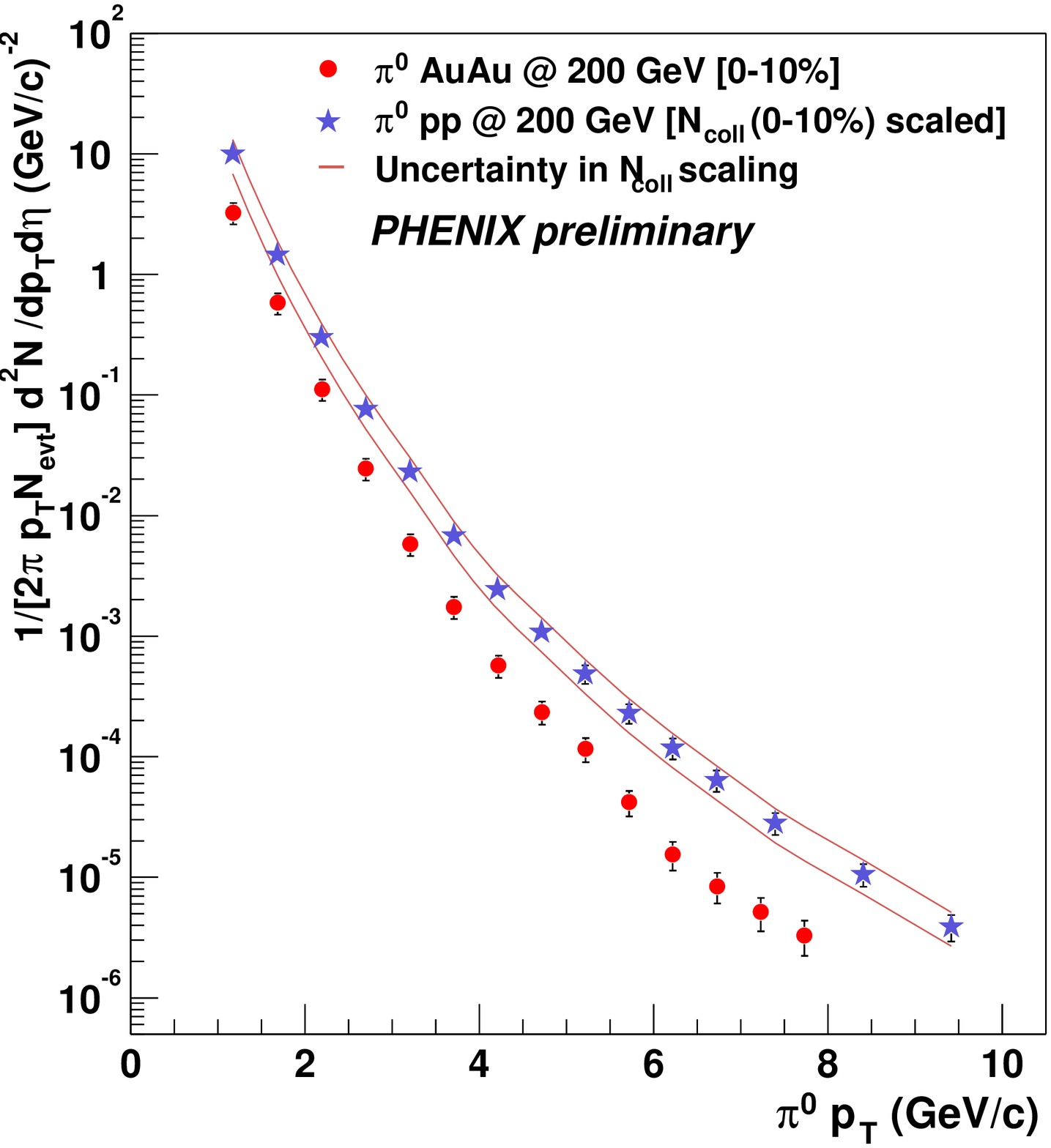}
\end{minipage}
\caption{Invariant $\pi^0$ yields measured in peripheral ({\it left}) and in central ({\it right}) 
$AuAu$ collisions ({\it circles}), compared to the $N_{coll}$ scaled $pp$ $\pi^0$ yields ({\it stars}) 
\protect\cite{enterria:hisa02}. The lines indicate the systematic errors in the scaled $pp$ 
yields due to the uncertainties in $N_{coll}$ and luminosity.}
\label{fig:pi0_spectra}
\end{figure}

Figure~\ref{fig:pi0_spectra} shows the resulting $\pi^0$ $p_T$ distributions 
measured for 70-80\% peripheral ({\it left}) and for 0-10\% central ({\it right})
$AuAu$ collisions, compared to the $\pi^0$ yield measured in $pp$

\hspace{-5mm}
%\noindent
collisions \cite{enterria:hisa02} scaled by the corresponding number of binary collisions given by 
a Glauber model calculation: $N_{coll}(70-80\%)$ = 12.3 $\pm$ 4.0, and $N_{coll}(0-10\%)$ = 975 $\pm$ 94 respectively.
Whereas $\pi^0$ production in peripheral reactions is consistent with the
point-like scaling expectation, the spectrum from the most central reactions is clearly 
suppressed (by a factor $\sim$2.5 at $p_T \sim$ 2 GeV/c 
increasing up to a factor $\sim$6 at $p_T \sim$ 8 GeV/c).
A complementary way to depict such a suppression is using the {\it nuclear modification factor}

\begin{equation}
R_{AA}(p_T)\,=\,\frac{d^2N^{\pi^0}_{AA}/d\eta dp_T}{\langle N_{coll}\rangle \, d^2N^{\pi^0}_{pp}/d\eta dp_T},
\label{eq:nucl_modif_factor}
\end{equation}

\begin{figure}[htb]
\begin{minipage}[t]{75mm}
  \includegraphics[scale=0.35]{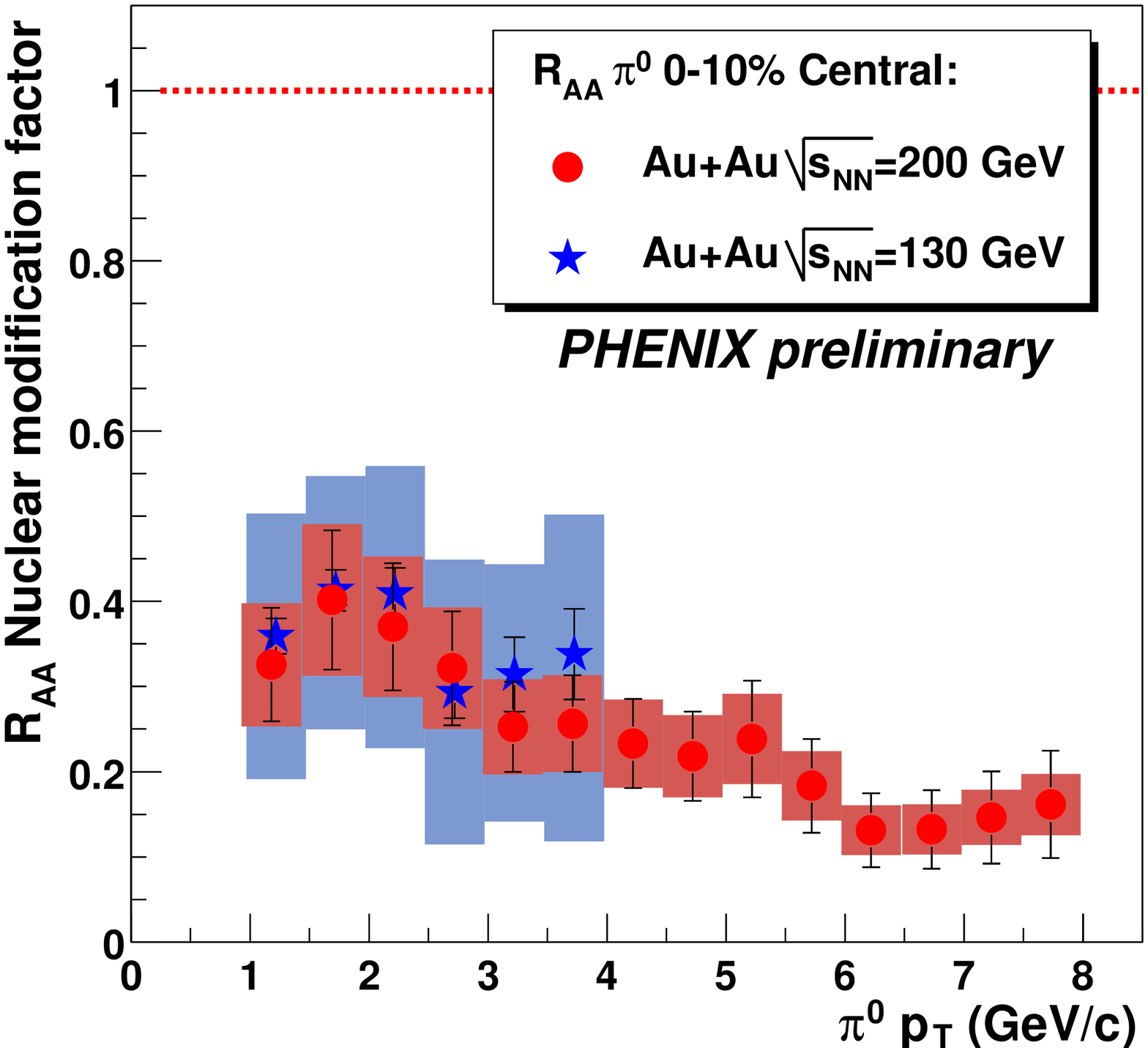}
\end{minipage}
%\hspace{\fill}
\begin{minipage}[t]{75mm}
  \includegraphics[scale=0.35]{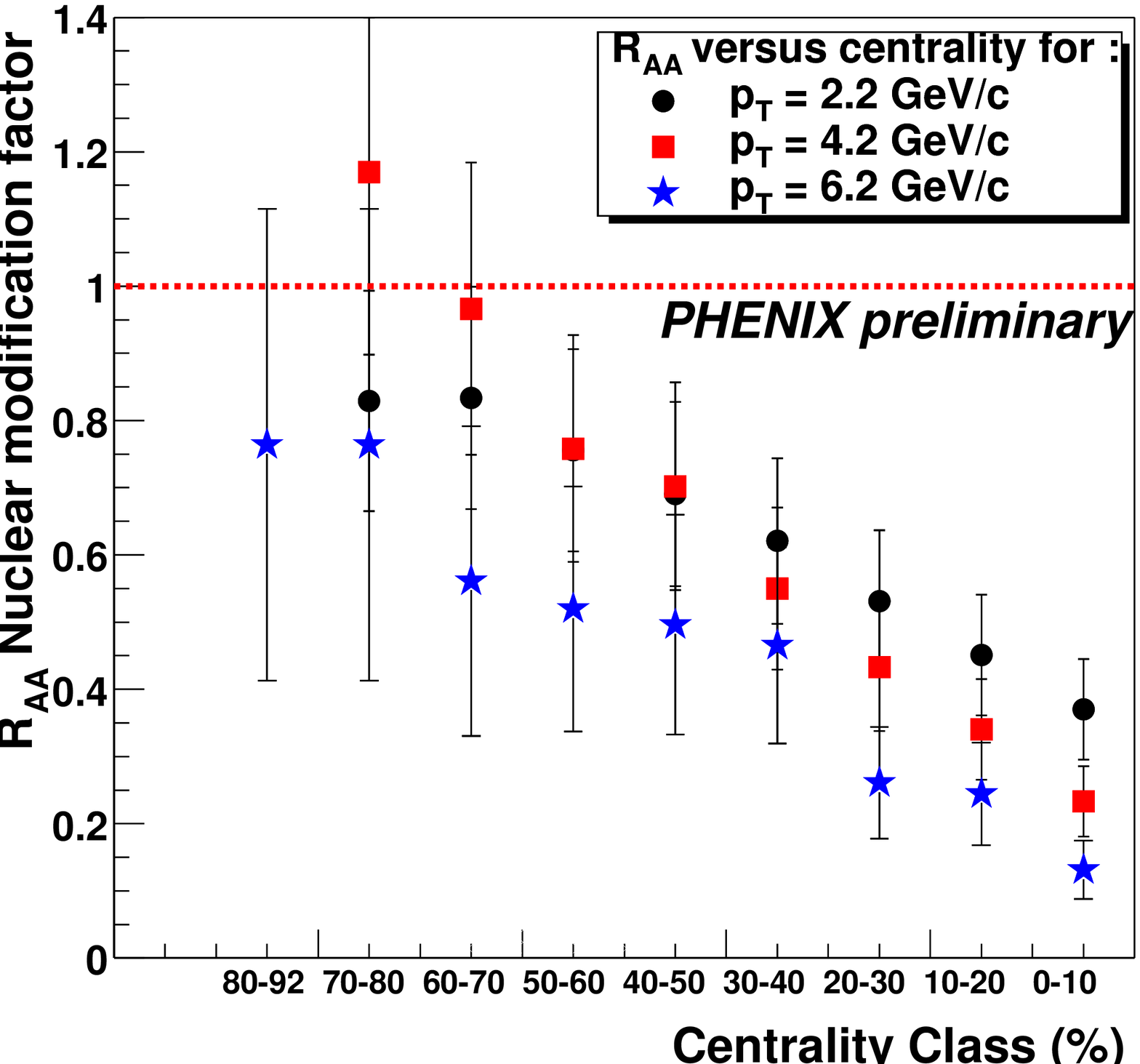}
\end{minipage}
\caption{{\it Left}: Nuclear modification factor %, Eq. (\protect\ref{eq:nucl_modif_factor}), 
for $\pi^0$ emitted in top 10\% central $AuAu$ collisions at 200 GeV 
%({\it red dots}), and 130 GeV ({\it blue dots}), and 17.3 GeV ({\it green squares}). 
({\it circles}), and 130 GeV ({\it stars}) (the bands %indicate the range of systematic 
show the syst. uncertainties). {\it Right}: $R_{AA}$ at 200 GeV as a function of the
reaction centrality for 3 fixed $p_T$ bins.}
\label{fig:R_AA}
\end{figure}

\hspace{-6mm}
%\noindent
which quantifies the deviation of the $\pi^0$ yield in $AA$ collisions with respect to the $pp$ behaviour 
(i.e. with respect to the absence of nuclear-medium effects) in terms of suppression or enhancement 
($R_{AA}$ smaller or larger than unity, respectively). Figure~\ref{fig:R_AA}{\it a} shows the 
$p_T$ dependence of $R_{AA}$ for $\pi^0$ emitted in 0-10\% central $AuAu$ reactions at 
two center-of-mass energies: 200 GeV ({\it circles}) and 130 GeV \cite{enterria:ppg003} ({\it stars}).
%(i) PHENIX 200 GeV ({\it red dots}), and (ii) PHENIX @ 130 GeV ({\it blue dots}) \cite{enterria:ppg003}.
%, and (iii) WA98 @ 17 GeV ({\it green squares}) \cite{enterria:wa98}.
Three interesting conclusions can be extracted from this plot above\footnote{For $p_T<$ 2 GeV/c, 
$R_{AA}$ is below unity in all cases since the bulk of particle production is due to soft processes which 
scale with the number of participant nucleons ($N_{part}$) rather than with $N_{coll}$.} $p_T$ = 2 GeV/c:\\

\begin{enumerate}
\item 
%\hspace{-6mm}
%\noindent
%{\it i)} 
The $R_{AA}$ values at 200 GeV and 130 GeV are compatible (within their corresponding systematic errors)
and clearly below unity: $R_{AA}$(2 GeV/c) $\sim$ 0.4 gradually decreasing down to $R_{AA}$(8 GeV/c) $\sim$ 0.16 for the
highest $\sqrt{s}$.
%\vspace{1.5mm}
\item 
%\noindent
%{\it ii)} 
The low $R_{AA}$ values measured at RHIC are clearly at variance with the high-$p_T$ hadron
production enhancement ($R_{AA}>$1) observed at CERN-SPS energies (WA98 data) \cite{enterria:wa98,enterria:saskia}:
No ``Cronin effect'' \cite{enterria:cronin} is observed.% at collider energies.
\item
%\vspace{1.5mm}
%\noindent
%{\it iii)} 
The trend of $R_{AA}$ below unity in the range $p_T= 2-8$ GeV/c (corresponding to parton fractional 
momenta $x=2p_T/\sqrt{s}\sim 0.02-0.1$) seems to be inconsistent with initial-state nuclear effects 
(``shadowing'' of the $Au$ parton distribution function). Indeed, shadowing is known \cite{enterria:eks98}
to decrease for larger $x$
% because, if that were the case, one would expect a similar 
%``anti-shadowing'' increase of $R_{AA}$ to start to set in for the highest $p_T$ bins ($x\sim$ 0.1 
(turning into ``anti-shadowing'' above $x\sim$ 0.1), whereas $R_{AA}$ consistenly diminishes for larger $p_T$ values.\\
\end{enumerate}

The same quenching is evident in the nuclear modification factor constructed not from the ratio of
$AuAu$ to $pp$ pions but from the ratio of central to peripheral $AuAu$ $\pi^0$ spectra (scaled by
their corresponding centrality-dependent $N_{coll}$). Such $R_{AA}$ (see fig. 5 of S. Mioduszewski 
contribution \cite{enterria:saskia}), -which have part of
the systematic errors on the spectra cancelled out-, are also well below one in central collisions. 
Finally, fig.~\ref{fig:R_AA}{\it b} shows the evolution of the $\pi^0$ suppression at three fixed
$p_T$ bins ($p_T$= 2.2, 4.2 and 6.2 GeV/c) as a function of the reaction centrality. The transition 
from the binary-scaling behaviour apparent in the peripheral region ($R_{AA}\sim$ 1, centrality class $>$ 70\%) 
to the highly suppressed central region is gradual and seems to occur over the 50-70\% centrality bin
(the systematic uncertainty in $R_{AA}$, not shown in this plot, is of the order of $\sim$20\%).

%in which part of the uncertainties cancel out (not shown here).
%Perturbative QCD calculations \cite{enterria:ina02} including initial-state effects (such as Cronin enhancement, 
%and nuclear shadowing), and modified fragmentation functions to take into account medium-induced energy loss are
%able to reproduce the suppression in central collisions with $p_T$-dependent energy losses 
%amounting up to $dE/dx\sim$ 1.0 GeV/fm. Peripheral collision spectra, on the other hand, are consistent 
%with $dE/dx\sim$ 0.

\section{Conclusions}
Transverse momentum spectra of neutral pions have been measured at mid-rapidity by the PHENIX EMCal 
up to $p_T$ = 8 GeV/c for 9 different centrality classes from 30$M$ Au+Au ``minimum bias'' events  
at $\sqrt{s_{\mbox{\tiny{\it{NN}}}}}$ = 200 GeV. The spectral shape and 
invariant yield of peripheral reactions are consistent with that of $pp$ collisions scaled
by the number of inelastic $NN$ collisions. Central yields, on the other hand, are
significantly lower than peripheral $AuAu$ and $pp$ binary-scaled extrapolations, in agreement with the
results found at $\sqrt{s_{\mbox{\tiny{\it{NN}}}}}$ = 130 GeV. The observed quenching sets in
over the 50-70\% centrality class and increases with $p_T$ and centrality, 
being as high as a factor 6 at $p_T$ = 8 GeV/c in the top 10\% central collisions.
For these collisions, the nuclear modification factor $R_{AA}$ is systematically
well below one and decreases gradually  with $p_T$, at variance with the common
initial-state ``phenomenology'': (i) no ``Cronin enhancement'' %, as found at lower energies, 
is observed, and (ii) the growing suppression with $p_T$ seems inconsistent with the decreasing
dependence of ``shadowing'' on $x$. Since initial-state nuclear effects seem not determinant, final-state
%Final-state ({\it partonic} rather than {\it hadronic}) high energy-density  
medium effects appear as a promising explanation for the 
%Since, any significant role of initial-state effects seems to be ruled out as an explanation for the 
observed high-$p_T$ pion deficit. 
%Since dense {\it hadronic} medium scenarios , the 
%only remaining explanation of these PHENIX data calls for high energy-density {\it partonic} medium formation.

% PHENIX central $AuAu$ high-$p_T$ results are thus in qualitative agreement with the
%expectations of parton energy loss in an opaque medium.
%This fact rules out strong shadowing as a possible explanation of the
%observed hadron deficit.


\begin{thebibliography}{9}
\bibitem{enterria:ppg003} K.~Adcox {\it et al.} [PHENIX Collaboration], Phys. Rev. Lett. 88 (2002) 022301.
\bibitem{enterria:ina02} I.~Sarcevic {\it et al.}, I.~Vitev {\it et al.}, and P.~Levai {\it et al.}, these Proceedings.
\bibitem{enterria:hisa02} H.~Torii [PHENIX Collaboration], these Proceedings.
\bibitem{enterria:wa98} K.~Reygers [WA98 Collaboration], nucl-ex/0202018.
\bibitem{enterria:saskia} S.~Mioduszewski [PHENIX Collaboration], these Proceedings.
\bibitem{enterria:cronin} %J.W.~Cronin {\it et al.}, Phys.~Rev.  D11 (1975) 3105.
D.~Antreasyan {\it et al.}, Phys. Rev. D19 (1979) 764.
\bibitem{enterria:eks98} K.~Eskola {\it et al.}, Phys. Lett. B532 (2002) 222. %and C.~Salgado, these Proceedings.
\end{thebibliography}
\end{document}